\documentclass[aps,twocolumn,showpacs]{revtex4}
\usepackage{graphicx}
\usepackage{bm}
\usepackage{amsmath}

\begin{document}

\title{Quantum Anomalous Hall Effect in Hg$_{1-y}$Mn$_{y}$Te Quantum Wells}
\author{Chao-Xing Liu$^{1,2}$, Xiao-Liang Qi$^{2}$,
Xi Dai$^{3}$, Zhong Fang$^{3}$ and Shou-Cheng Zhang$^{2}$}

\affiliation{$^1$ Center for Advanced Study, Tsinghua
University,Beijing, 100084, China} \affiliation{$^2$ Department of
Physics, McCullough Building, Stanford University, Stanford, CA
94305-4045} \affiliation{$^{3}$Institute of Physics, Chinese Academy of Sciences
Beijing, 100080, China}

\begin{abstract}
The quantum Hall effect is usually observed when the two-dimensional
electron gas is subjected to an external magnetic field, so that
their quantum states form Landau levels. In this work we predict
that a new phenomenon, the quantum anomalous Hall effect, can be
realized in Hg$_{1-y}$Mn$_{y}$Te quantum wells, without the external
magnetic field and the associated Landau levels. This effect arises
purely from the spin polarization of the $Mn$ atoms, and the
quantized Hall conductance is predicted for a range of quantum well
thickness and the concentration of the $Mn$ atoms. This effect
enables dissipationless charge current in spintronics devices.
\end{abstract}


\maketitle

When the two-dimensional electron gas (2DEG) is subjected to a high
magnetic field, electronic states form Landau levels. In the regime
where the temperature is low compared to the spacing between the
Landau levels, quantized Hall conductance can be observed. In the
quantum Hall (QH) regime, the electric current flows
uni-directionally along the edge of the sample. Since
back-scattering is absent, the edge current flows without any
dissipation. The breaking of the time reversal symmetry is a
necessary condition for the Hall effect. However, an external
magnetic field is not required. Soon after the observation of the
Hall effect, Edwin Hall also observed the anomalous Hall
effect\cite{Chien1980}, where an additional Hall resistance arises
from the spin-orbit interaction between the electric current and the
magnetic moments. In the extreme case, an anomalous Hall effect can
occur without the external magnetic field as long as the system
breaks time-reversal symmetry spontaneously. Given the experimental
observation of the QH effect, it is natural to ask whether the
anomalous Hall effect can also be quantized without the external
magnetic field and the associated Landau levels. Such a question is
not only of great academic interests, but also has important
practical implications. Realizing dissipationless charge current
through the quantum anomalous Hall (QAH) effect without an external
magnetic field could enable a new generation of quantum electronic
devices.

Some years ago, Haldane\cite{Haldane1988} constructed a theoretical
toy model to show that the QH effect is in principle possible
without Landau levels. The model describes the hopping of spinless
fermions on a honeycomb lattice, where the next-nearest neighbor
hopping has complex amplitudes. Even though the complex hopping
amplitude breaks the time reversal symmetry, it does not break the
lattice translation symmetry since the net magnetic flux vanishes
inside an unit cell. Haldane's model was extended to include
localization physics in Ref.\cite{onoda2003}. Unfortunately, this
model is mostly academic, and can not be realized in the recently
discovered graphene system. Later, the work on intrinsic anomalous
Hall effect of ferromagnetic semiconductor in the metallic
regime\cite{Jungwirth2002, Fang2003,Haldane2004} shows the relation
between the Berry's phase and Hall conductance. Qi, Wu and
Zhang\cite{Qi2006} constructed a tight-binding model of electrons
spin-orbit coupled to polarized magnetic moments, and showed that
the Hall conductance can be quantized in appropriate parameter
regimes. This model breaks the time reversal symmetry due to
magnetic moments rather than Landau levels, and provides another
example of the QAH effect. More recently, a closely related
topological phenomenon known as quantum spin Hall (QSH) effect has
been theoretically predicted and experimentally observed in HgTe
quantum wells\cite{Bernevig2006,MarkusKonig2007}. In this work, we
show that when the HgTe quantum wells are doped with the magnetic Mn
atoms, the QAH effect can be realized within an experimentally
accessible parameter regime. We propose an experiment to demonstrate
that the quantized Hall conductance indeed arises from the magnetic
moments rather than Landau levels.

As a starting point, we first briefly review the physics of QSH
effect in HgTe quantum wells. HgTe has an inverted band structure,
where the p-type $\Gamma_8$ band has higher energy compared to the
s-type $\Gamma_6$ band at the $\Gamma$ point. For HgTe/CdTe
quantum wells, there exists a topological quantum phase transition
across some critical well thickness $d_c$ where the band structure
changes from the normal to the inverted character. The novel QSH
effect occurs in the inverted regime $d>d_c$\cite{Bernevig2006}.
In order to describe the physics near $d_c$, an effective
four-band model is introduced as
\begin{eqnarray}
    H_0(k)=\left(
    \begin{array}{cc}
        h_+(k)&0\\
        0&h_-(k)
    \end{array}
    \right),
    \label{Ham_0}
\end{eqnarray}
where
\begin{eqnarray}
h_+(k)&=&\left(\begin{array}{cc}\epsilon_{k}+\mathcal{M}(k)&Ak_+\\Ak_-&\epsilon_{k}
-\mathcal{M}(k)\end{array}\right). \label{h_0}
\end{eqnarray}
and $h_-(k)=h_+^*(-k)$ as is required by time-reversal symmetry.
$\epsilon_{k}$ and $\mathcal{M}(k)$ can be in general expanded as
\begin{eqnarray}
    \epsilon_{k}=C_0+C_2k^2,\qquad
    \mathcal{M}(k)=M_0+M_2k^2.
    \label{coeffi1}
\end{eqnarray}
This effective model is expressed in the subspace containing the
states $|E1,\pm\rangle$ and $|H1,\pm\rangle$, where
$|E1,\pm\rangle$ is a superposition of
$|\Gamma_6,\pm\frac{1}{2}\rangle$ and
$|\Gamma_8,\pm\frac{1}{2}\rangle$ states, while $|H1,\pm\rangle$
is formed by $|\Gamma_8,\pm\frac{3}{2}\rangle$ states. Here $\pm$
denote the two spin states which are degenerate due to the Kramers
theorem when the time reversal symmetry is preserved. The diagonal
block $h_\pm(k)$ describes a Dirac model in $2+1$ dimensions,
which at half filling carries a Hall conductance of $\pm e^2/h$,
respectively\cite{Haldane1988,Qi2006,Bernevig2006}. Thus the net
Hall conductance of the inverted quantum well system vanishes,
while the spin Hall conductance, defined as the difference between
the two blocks, is still non-zero. Therefore the QSH effect can be
viewed as two copies of the QAH effects, with the opposite quanta
of Hall conductances.

When the time reversal symmetry is broken, the two spin blocks are
no longer related, and their charge Hall conductances no longer
cancel exactly. {\it The key idea of this work is to identify the
parameter space where one spin block is in the normal regime,
while the other spin block is in the inverted regime.} The normal
regime gives a topologically trivial insulator with vanishing Hall
conductance, while the inverted regimes gives a topologically
non-trivial insulator with one quantum unit of the Hall
conductance; therefore, the whole system becomes a QAH state. Now
we return to the four band effective Hamiltonian (\ref{Ham_0}) and
address what kind of term can induce the QAH effect. To describe
the spin splitting induced by the magnetization, a
phenomenological term is introduced as
\begin{eqnarray}
    H_{s}=\left(\begin{array}{cccc}G_E&0&0&0\\0&
        G_H&0&0\\0&0&-G_E&0\\0&0&0&-G_H\end{array}\right).
    \label{H_ss}
\end{eqnarray}
where the spin splitting is $2G_E$ for the $|E1,\pm\rangle$ band and
$2G_H$ for the $|H1,\pm\rangle$ band. Then the energy gap is given
by $E_g^+=2M_0+G_E-G_H$ for the up spin block while
$E_g^-=2M_0-G_E+G_H$ for the down spin block. In order to obtain the
QAH effect, we require that (1) the state with one kind of spin is
in the inverted regime while the other goes into the normal regime,
namely $E^+_gE^-_g<0$; (2) the entire system is still in the
insulating phase with a full bulk gap, which requires that the
states $|E1,+\rangle$ ($|E1,-\rangle$) and $|H1,-\rangle$
($|H1,+\rangle$) do not cross each other, leading to the condition
$(2M_0+G_E+G_H)(2M_0-G_E-G_H)>0$. Combining the two conditions
discussed above, we arrive at the condition $G_EG_H<0$, which
requires that the spin splittings for $|E1,\pm \rangle$ and
$|H1,\pm\rangle$ must have the opposite sign. This condition is
illustrated in Fig. 1(a). We can also understand the physics from
the edge state picture (Fig 1 (b)). On the boundary of a QSH
insulator there are counter-propagating edge states carrying
opposite spin. When the spin splitting term increases, the spin down
edge states penetrate much deeper into the bulk due to the
decreasing gap, and eventually disappear, leaving only the spin up
state bound more strongly to the edge. Thus the system has only spin
up edge states and transforms from the QSH state to the QAH state,
as illustrated in Fig 1 (b).

%In Fig 1 (a), we start with the QSH state in
%the $d>d_c$ regime, and then switch on the spin splitting term
%(\ref{H_ss}), assuming $G_E<0$ and $G_H>0$. With increasing $|G_E|$
%and $|G_H|$, the spin down pair closes its gap and approaches the
%normal regime (the blue dashed line cross each other in Fig 1 (a)),
%while spin up pair remains in the inverted regime. We can also
%understand the physics from the edge state picture (Fig 1 (b)). When
%the spin splitting term increases, the spin down edge states
%penetrate much deeper into the bulk, due to the decreasing gap, and
%eventually disappear, leaving only the spin up state bounded more
%strongly to the edge. The system has only spin up edge states and
%transforms from the QSH state to the QAH state (the last picture in
%Fig 1 (b)). For the opposite case $G_EG_H>0$, from Fig 1 (c), we
%find that $|H1,-\rangle$ and $|E1,+\rangle$ cross each other without
%opening a gap because there is no coupling between $|H1,-\rangle$
%and $|E1,+\rangle$. Therefore the system enters a metallic phase
%after the crossing.

\begin{figure}
    \begin{center}
        \includegraphics[width=3in]{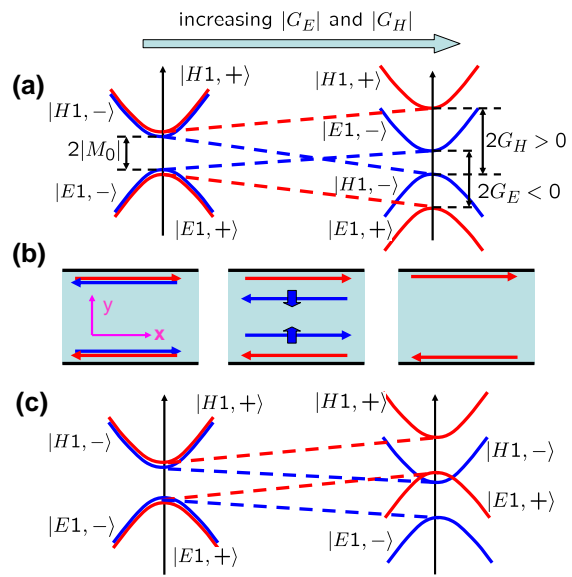}
    \end{center}
    \caption{(Color online) Evolution of band structure and edge states upon increasing the
    spin splitting. For (a) $G_E<0$ and $G_H>0$, the spin down states $|E1,-\rangle$ and
    $|H1,-\rangle$ in the same block of Hamiltonian (\ref{Ham_0})
    touch each other and then enters the normal regime. But for (c) $G_E>0$ and
    $G_H>0$, gap closing occurs between $|E1,+\rangle$ and $|H1,-\rangle$ bands, which belongs to different blocks
    of Hamiltonian, and thus will cross each other without opening a gap. In (b), we show the behavior of the
    edge states during the level crossing in the case of (a).}
    \label{fig:schematic_edge}
\end{figure}

We have therefore identified a key condition $G_EG_H<0$, for the
QAH effect. We may ask whether or not it is true in the realistic
material. Fortunately, in the Cd$_x$Mn$_y$Hg$_{1-x-y}$Te
/Mn$_y$Hg$_{1-y}$Te quantum well, the sp-d exchange coupling
indeed gives the opposite signs for $G_E$ and $G_H$, a fact which
is well established in the
literature\cite{Bhattacharjee1983,Furdyna1988}. From a standard
perturbative treatment of the eight band Kane
model\cite{Winkler2003,Bernevig2006}, we find for a spin
polarization perpendicular to the quantum well plane, the
coefficients $G_E$, $G_H$ in the four-band effective model
(\ref{H_ss}) can be expressed as
\begin{eqnarray}
G_E=-(3AF_1+BF_4),\ \ G_H=-3B, \label{paraGH}
\end{eqnarray}
in which $A,B$ are given by\cite{Furdyna1988,Novik2005}
\begin{eqnarray}
A=\frac{1}{6}N_0\alpha y\langle \mathcal{S}\rangle,\ \
B=\frac{1}{6}N_0\beta y\langle \mathcal{S}\rangle,
\end{eqnarray}
and $F_1$, $F_4$ are the amplitudes of
$|\Gamma_6,\pm\frac{1}{2}\rangle$ and
$|\Gamma_8,\pm\frac{1}{2}\rangle$ components in the state
$|E_1,\pm\rangle$, respectively. Here $N_0$ is the number of unit
cells per unit volume and $\langle \mathcal{S}\rangle$ is the spin
polarization of Mn out of quantum well plane. $\alpha$ and $\beta$
describe the sp-d exchange coupling strength for s-band and p-band
electron respectively, where the signs and magnitudes are crucial
for the relative sign of $G_E$ and $G_H$. For ${\rm
Hg_{1-y}Mn_yTe}$, these parameters are determined to be
$N_0\alpha=0.4eV$ and
$N_0\beta=-0.6eV$\cite{Novik2005,Furdyna1988}, leading to the
opposite signs for $G_E$ and $G_H$. Combined with the previous
analysis about the Hamiltonian (\ref{H_ss}), we conclude that QAH
effect can occur in the ${\rm
Cd_xMn_yHg_{1-x-y}Te/Mn_yHg_{1-y}Te}$ quantum well, as long as the
${\rm Mn}$ magnetization $\left\langle \mathcal{S}\right\rangle$
is large enough and perpendicular to the quantum well plane.

%\begin{figure}
%    \begin{center}
%        \includegraphics[width=3in]{d_Eng.EPS}
%    \end{center}
%    \caption{ (a) The energy levels for $|E1,\pm\rangle$ and $|H1,\pm\rangle$
%    are plotted as a function of the quantum well thickness.
%    Two crossing points (A and B) are
%    labelled in the figure. (b) energy dispersion at the thickness $d_{c1}=7.25nm$; (c)
%    energy dispersion at the thickness $d_{c2}=9.4nm$. Here the magnetization of the Mn atom
%    is chosen to be $\langle \mathcal{S}\rangle=2$.
%}
%    \label{fig:d_Eng}
%\end{figure}

\begin{figure}
    \begin{center}
    \includegraphics[width=3.2in]{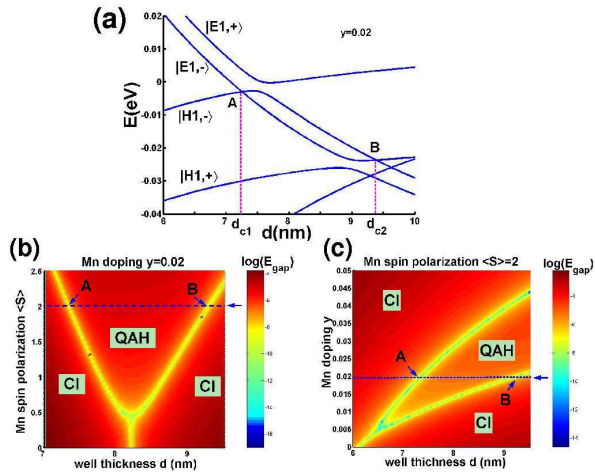}
    \end{center}
    \caption{(a) The energy levels for $|E1,\pm\rangle$ and $|H1,\pm\rangle$
    are plotted as a function of the quantum well thickness. Two crossing points (A and B) are
    labelled in the figure. The energy gap ($log(E_{gap})$ used here) is
    plotted as a function of the well thickness $d$ versus the $Mn$ magnetic moment $\langle S\rangle$ in
    (b), versus the $Mn$ doping concentration $y$ in (c). Dashed blue line in (b) or (c) refers to the line along
    which (a) is plotted. The points 'A' and 'B' correspond to the two Dirac type
    crossing points. Two different phases, conventional insulator
    (CI) with $\sigma_H=0$ and QAH state with $\sigma_H=-e^2/h$, are separated by the gap closing line in the figures. }
    \label{fig:phadia_Ck}
\end{figure}

%The energy gap ($log(E_{gap})$ used here) is
%    plotted as a function of the well thickness $d$ versus the $Mn$ magnetic moment $\langle S\rangle$ in
%    (a), versus the $Mn$ doping concentration $y$ in (b). Dashed blue line in (a) or (b) refers to the line along
%    which Fig 2 is plotted. The points 'A' and 'B' correspond to the two Dirac type
%    crossing points in Fig 2. Two different phases, conventional insulator
%    (CI) with $\sigma_H=0$
%    and QAH state with $\sigma_H=-e^2/h$, are separated by the gap closing line in the figures.

Though the analysis above already gives a clear explanation to the
physical mechanism of QAH effect, to obtain more quantitative
predictions we perform a more realistic calculation of the
electronic structure based on the eight band Kane
model\cite{Novik2005,Winkler2003}. Our Hamiltonian takes into
account the exchange term and bulk inversion asymmetry (BIA)
terms\cite{Winkler2003}, in addition to the terms included in the
original Kane model. In Fig. 2 (a), the energy spectrum of the
lowest subbands $|E1,\pm\rangle$ and $|H1,\pm\rangle$ at the
$\Gamma$ point is plotted as a function of quantum well thickness
$d$. At a critical thickness $d_{c1}=7.25{\rm nm}$, a level
crossing `A' occurs between the $|E1,-\rangle$ and the
$|H1,-\rangle$ bands. This is a Dirac-type level crossing, which
corresponds to a topological quatnum phase transition, across
which the Hall conductance jumps by $-e^2/h$. Since the system at
$d<d_{c1}$ always remains gapped when the $Mn$ magnetization is
adiabatically turned on, we know $\sigma_H=0$ for $d<d_{c1}$ and
$\sigma_H=-e^2/h$ for $d>d_{c1}$. The same analysis applies to the
other critical thickness $d_{c2}=9.4{\rm nm}$ (`B' in Fig 2 (a)),
where the level crossing occurs between $|E1,+\rangle$ and
$|H1,+\rangle$ bands and the Hall conductance returns to $0$.
Therefore, for parameters $\left\langle
\mathcal{S}\right\rangle=2$ and $y=0.02$, QAH effect appears in a
quantum well thickness range $d_{c1}<d<d_{c2}$. The same
calculation can be carried for different spin polarizations
$\left\langle \mathcal{S}\right\rangle$ or $y$ to determine the
whole phase diagram with three tunable parameters: quantum well
thickness $d$, spin polarization $\langle \mathcal{S}\rangle$ and
$Mn$ fraction $y$. In Fig. 2 (b) and (c), the gap sizes in the
$d$-$\langle\mathcal{S}\rangle$ plane and the $d$-$y$ plane are
plotted, respectively. The line with bright color shows the phase
boundary with vanishing gap. %The blue dashed line in Fig 2 (b) and
%(c) both correspond to the line along which Fig 2 is drawn.
%Therefore, from the analysis in Fig 2 we identify the regime between
%the two gap closing line as the QAH regime.
At zero $\langle \mathcal{S}\rangle$ or zero $y$ limit, the
time-reversal invariance is recovered and the critical line
terminates at the transition point between trivial and QSH
insulators\cite{Bernevig2006}. An important feature of the phase
diagram is that a minimal $\langle \mathcal{S}\rangle$ (for fixed
$y$) or minimal $y$ (for fixed $\langle \mathcal{S}\rangle$) is
required to obtain QAH phase. For example, for $y=0.02$ we need
$\langle \mathcal{S}\rangle>0.5$ for the QAH phase. This is a
consequence of bulk inversion asymmetry (BIA).

We now discuss the experimental observation of the QAH effect in the
Hg$_{1-y}$Mn$_{y}$Te system. The main difficulty lies in the fact
that Hg$_{1-y}$Mn$_{y}$Te is paramagnetic rather than ferromagnetic.
Thus the spin polarization is determined by the Curie-Weiss formula:
\begin{eqnarray}
    \langle\mathcal{S}\rangle&=&-S_0B_{5/2}\left(\frac{5g_{Mn}\mu_BB}{2k_B(T+T_0)}\right)
    \label{paraS}
\end{eqnarray}
where $S_0=5/2$ and $B_{5/2}$ is the Brillioun
function\cite{Winkler2003,Novik2005}, and $T_0>0$ stands for a
weak antiferromagnetic coupling between $Mn$ spins. Hence, we need
to apply a small magnetic field to polarize the Mn moments and
generate the QAH effect. In fact, experiments carried out at the
University of W\"{u}rzburg has already shown quantized Hall
conductance for magnetic fields as small as
$1T$\cite{Buhmann2002}. However, since even such a small magnetic
field still has an orbital effect, this experiment by itself
cannot prove the existence of the QAH, which arises purely from
the magnetic moments of ${\rm Mn}$. To solve this problem, we
propose two different ways to polarize Mn spins without an orbital
magnetic field. The first approach is to apply a low-frequency
polarized infrared light to provide the angular momentum required
for aligning the Mn moments. Such a photo-induced Mn magnetization
has been observed experimentally both in ${\rm Hg_{1-x}Mn_{x}Te}$
and ${\rm Cd_{1-x}Mn_{x}Te}$\cite{Krenn1985,Awschalom1987}. Though
this proposal has the advantage of realizing this effect in a
steady state, another approach, the time-resolved Hall
measurement, can provide a more dramatic demonstration of the QAH
effect. Thus in the following we will focus on the time-resolved
Hall measurement proposal. Firstly, a static magnetic field is
applied, which acts on the itinerant electrons through three
terms: the orbital effect $H_{\rm orb}$, the Zeeman term $H_{\rm
Z}$ and the exchange term $H_{\rm ex}$. In $H_{\rm ex}$, the ${\rm
Mn}$ spin polarization is determined by Eq. (\ref{paraS}).
Secondly, at time $t_0$ the magnetic field is switched off within
a time scale $\tau_B$. As a result, the ${\rm Mn}$ spin
polarization will decay to zero in a spin-relaxation time
$\tau_s$. However, if $\tau_B\ll \tau_s$, there is a time window
$t_0+\tau_B<t\ll t_0+\tau_s$ when the orbital and Zeeman effect of
the magnetic field already disappeared, but the ${\rm Mn}$ spin
polarization remains similar to the value before magnetic field is
removed. Consequently, within this time range there is no
conventional QH based on Landau levels, so that the pure QAH
effect can be observed, if the ${\rm Mn}$ spin polarization stays
in the correct range.

\begin{figure}
    \begin{center}
        \includegraphics[width=3in]{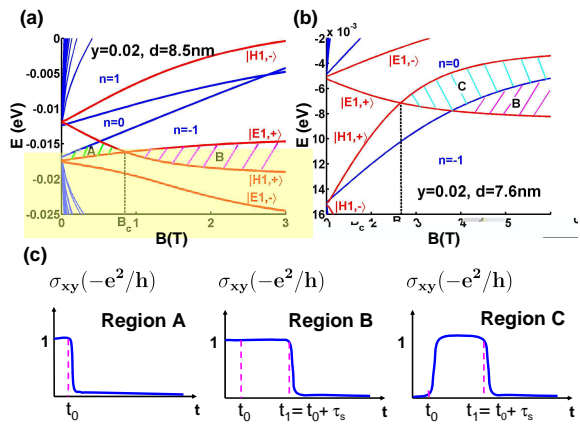}
    \end{center}
    \caption{ Landau level spectra are plotted as a function of magnetic field (blue line)
    for two different thicknesses (a) $d=8.5nm$ and (b) $d=7.6nm$, taking into account exchange term, Zeeman
    splitting and orbital effect. The corresponding red lines show the
    band edge of the sub-bands $|H1,\pm\rangle$ and $|E1,\pm\rangle$ when the magnetic field has been turned
    off but the Mn spin polarization remains.
    Without an external magnetic field, the system is in a QSH insulator phase for (a),
    and trivial insulator phase for (b).
    Figure (c) shows schematically the predicted Hall conductances for the three different regions A, B and C
    defined by shadows in figure (a) and (b).}
    \label{fig:EngSpe_Com}
\end{figure}

To illustrate this proposal more clearly, we compare the band
structure with and without Landau levels. In Fig 4 (a) and (b), the
energy spectra is plotted as a function of the magnetic field for
different thicknesses. The blue lines take into account the orbital
magnetic field term, the exchange and the Zeeman terms,
corresponding to the spectra at time $t<\tau_B$, (setting $t_0=0$).
In contrast, the red lines are the conduction and valence band edges
of the system at time $\tau_B<t\ll\tau_s$, after the orbital and
Zeeman effects of the magnetic field are switched off but the ${\rm
Mn}$ spin polarization remains the same. Since the energy spectrum
is dispersive when the orbital effect of magnetic field is absent,
the system is insulating only when the fermi energy lies between the
two middle red lines, {\em i.e.}, between $|E1,-\rangle$ and
$|H1,-\rangle$ for Fig. 4 (a), or between $|E1,+\rangle$ and
$|H1,+\rangle$ for Fig. 4 (b). Depending on the initial values of
fermi energy $E_F$ and magnetic field $B$, three different phenomena
can be observed in the proposed experiment. (i) When $(E_F,B)$ is in
Region (A) of Fig. 4 (a), the system has Hall conductance
$\sigma_H=-e^2/h$ in the static magnetic field, and enters a trivial
insulator phase after magnetic field is switched off. Consequently,
the Hall conductance will drop to zero once the magnetic field is
removed. (ii) When $(E_F,B)$ is in Region (B) of Fig. 4 (a) and (b),
the system has Hall conductance $\sigma_H=-e^2/h$ for static field
and enters a QAH phase with the same Hall conductance after the
magnetic field is switched off. Thus the Hall conductance will
remain on the $-e^2/h$ plateau for a time $\sim \tau_s$ after
magnetic field is turned off. (iii) When $(E_F,B)$ is in Region (C)
of Fig. 4 (b), the system has vanishing Hall conductance in static
field, but enters a QAH phase with $\sigma_H=-e^2/h$ after the
magnetic field is switched off. Consequently, we will observe the
dramatic appearance of a ``pulse" of the Hall conductance in a
time-scale $\sim \tau_s$ even though the system is in the
$\sigma_H=0$ Hall plateau under a static field! Observation of this
phenomenon gives the conclusive demonstration of the QAH effect. Due
to the topological distinction between $\sigma_H=0$ and
$\sigma_H=-e^2/h$ states, the transition between them is always
sharp at low temperature, even though the ${\rm Mn}$ magnetization
changes continuously. Consequently, the time-dependence of
$\sigma_{xx}$ and $\sigma_{xy}$ in this proposal should show the
same critical behavior as the usual ``plateau transition" in the QH
effect.

Finally, we estimate the experimental conditions required to
realize this proposal. First of all, both of the switch-off time
$\tau_B$ and the time resolution $\delta t$\cite{Lu2003} of the
Hall measurement should be shorter than Mn spin relaxation time
$\tau_s$, which is of the order $10-100\mu
s$\cite{Scherbakov2000,Witowskia1995}. Further, the magnetic field
should be turned off slowly enough for electrons to stay in the
instantaneous ground state during the switch-off operation
(adiabatic condition), which requires $\tau_B\gg \hbar/E_g$, where
$E_g$ is the energy gap and $\hbar/E_g\sim 10^{-12}s$ in the
present system. Therefore, a magnetic field of several Tesla needs
to be switched off in a time scale of $10^{-12}s\ll\tau_B\ll
10^{-4}s$, and the time-resolution of transport measurement should
satisfy $\delta t\ll 10^{-4}s$.

%Once QAH effect is established, one can achieve the static QAH state
%simply by using a hybrid ferromagnet-semiconductor
%structure\cite{Prinz1990}, where a small, static magnetic field by
%itself is not sufficient to generate the conventional QH effect, but
%it can polarize the Mn moments enough to cause the QAH effect. For
%example, a hybrid ferromagnetic-semiconductor structure has been
%realized with a magnetic layer deposited on the top of the
%semiconductor structure, where the static magnet field can be of the
%order $1T$\cite{Prinz1990}, sufficient to bring the
%Hg$_{1-y}$Mn$_{y}$Te system into the QAH regime.

We have benefited greatly from the close interaction with our
experimental colleagues at the University of W\"{u}rzburg, in
particular, H. Buhmann, M. K\"{o}nig and L. Molenkamp; without their
generous sharing of the experimental data and their insights on the
experimental feasibility this work would not have been possible. The
authors also would like to thank K. Chang, T. Hughes, R. B. Liu, S.
Q. Shen, J. Wang, F. Ye and B. F. Zhu for helpful discussions. This
work is supported by the NSF under grant numbers DMR-0342832 and the
US Department of Energy, Office of Basic Energy Sciences under
contract DE-AC03-76SF00515. XD and ZF acknowledge support from the
NSF and National Basic Research (973) Program of China
(No.2007CB925000). CXL acknowledges the support of China Scholarship
Council, NSF of China (Grant No.10774086, 10574076), and the Program
of Basic Research Development of China (Grant No. 2006CB921500).

%\bibliography{QAH_PRL}

\end{document}